\documentclass[twocolumn,prl,aps,superscriptaddress
,showpacs,
nofootinbib,preprintnumbers,floats,floatfix]{revtex4}

\usepackage{amsmath}
\usepackage{amssymb}
\usepackage{latexsym}
\usepackage{graphicx}
\usepackage{hyperref}
\usepackage{bm}

\begin{document}

\title{Cosmic microwave background anomalies in an open universe}

\author{Andrew R.~Liddle} 
\affiliation{Institute for Astronomy, University of Edinburgh, Blackford Hill, Edinburgh EH9 3HJ, United Kingdom} 
\author{Marina Cort\^es} 
\affiliation{Institute for Astronomy, University of Edinburgh, Blackford Hill, Edinburgh EH9 3HJ, United Kingdom} 
\affiliation{Centro de Astronomia e Astrof'sica da Universidade de Lisboa, Faculdade de Ci\^encias, Ed. C8, Campo Grande, 1769-016 Lisboa, Portugal}
\date{\today}

\begin{abstract}
We argue that the observed large-scale cosmic microwave anomalies, discovered by {\it WMAP} and confirmed by the {\it Planck} satellite, are most naturally explained in the context of a marginally-open universe. Particular focus is placed on the dipole power asymmetry, via an open universe implementation of the large-scale gradient mechanism of Erickcek et al. Open inflation models, which are motivated by the string landscape and which can excite `super-curvature' perturbation modes, can explain the presence of a very-large-scale perturbation that leads to a dipole modulation of the power spectrum measured by a typical observer. We provide a specific implementation of the scenario which appears compatible with all existing constraints.
\end{abstract}


\maketitle

\section{Introduction}

The confirmation by the {\it Planck} satellite \cite{planckanom} of puzzling anomalies in the large-scale cosmic microwave sky found in data from the {\it WMAP} mission \cite{WMAP9} has led to renewed interest in the possibility that they may have an underlying physical explanation. The various anomalies include alignments of the lowest cosmic microwave background (CMB) multipoles \cite{TOHevil}, a large cold spot \cite{coldspot}, a general power deficit for multipoles below 40 \cite{planckanom}, and a dipole power asymmetry extending to small angular scales \cite{asym,planckanom}. While none of these anomalies individually carries great significance \cite{WMAPanom}, they are collectively troubling despite the success of the standard cosmological model in fitting the precision data at higher multipoles and deserve to be taken seriously; see e.g.\ Refs.~\cite{erickcek,DJKC}. In keeping with recent papers on the topic, e.g.\ Refs.~\cite{lyth,newpapers,NBF}, we will focus on the dipole power asymmetry as it is the best quantified, though a truly compelling explanation ought to simultaneously explain several anomalies. Following Refs.~\cite{WMAPanom,planckanom} we refer specifically to the case of a dipolar modulation of the power spectrum, distinguishing this from other forms of hemispherical asymmetry or more general power spectrum modulation, say via bipolar spherical harmonics beyond the dipole contribution.

If there is an underlying physical cause for the anomalies, the implication is that there is a new large physical scale relevant to cosmology, beyond which the extrapolation of the standard $\Lambda$CDM cosmology breaks down. There are at least three candidates for such a scale; a topological identification scale for the Universe, the scale corresponding to the beginning of cosmological inflation (or a sharp exit of our region from an eternally inflating phase), and the curvature scale of a non-flat universe.

It ought to be possible to explain the anomalies with cosmic topology, as it can naturally break isotropy. However searches in data have found no evidence \cite{planckxxvi}, and it is not yet clear whether it would be possible to explain the anomalies without generating other signals that should have been detected, such as circles-in-the-sky. If inflation started not long before our observable Universe left the inflationary horizon, we should expect large-scale perturbations which could readily affect our Universe on the largest scales, but it is not clear that such effects are calculable. We therefore wish to focus this article on the third possibility, that the new scale is the curvature scale.

Present constraints from data, particularly {\it Planck} \cite{planckxvi}, assure us that the Universe is within a percent or so of spatial flatness. In an open universe, this corresponds to a limit that the comoving curvature scale
\begin{equation}
r_{{\rm curv}} = \frac{a^{-1} H^{-1}}{\sqrt{|1-\Omega_{\rm total}}|}\,,
\end{equation}
where $\Omega_{{\rm total}}$ is the density parameter including any dark energy component, is at least ten times the Hubble radius $H^{-1}/a$. This however is only about three times the size of the observable Universe, insufficient to guarantee no observable effects associated to the curvature scale.

It is tempting to think that the limit on curvature is indicating that the Universe is flat on average. However, there has been renewed motivation to consider marginally open universes in the context of string landscape cosmology \cite{landscape}, where it is quite plausible that our region of the Universe arose from decay of an meta-stable false vacuum state via bubble nucleation. Such tunnellings, via the Coleman--de Luccia instanton \cite{coldel}, have long been known to create effective open universes, and moreover the difficulty in obtaining very large amounts of inflation within the landscape \cite{freivogel} suggests that the subsequent inflation within the bubble may not last so long as to establish a universe indistinguishably close to flat \cite{openinf,BGT,LM}. Extensive arguments for searching for curvature effects under these motivations have, for instance, been made by Yamauchi et al.~\cite{openland}.

\section{Modulation by large-scale perturbations}

The observed power asymmetry in the CMB sky can be modelled as a dipole modulation of the power spectrum of an otherwise statistically-isotropic sky \cite{gordon2}
\begin{equation}\label{defA}
\Delta T(\hat{{\bf n}}) = \left( 1+ A \hat{{\bf p}}\cdot\hat{{\bf n}} \right) \Delta T_{{\rm iso}}(\hat{{\bf n}})\,,
\end{equation}
where $\hat{{\bf n}}$ and $\hat{{\bf p}}$ are unit vectors in a sky direction and in the dipole modulation direction respectively. The modulation amplitude $A$ is measured in the CMB to be $0.07 \pm 0.02$ in maps smoothed at 5 degrees \cite{planckanom}.

A mechanism to explain such a modulation, introduced by Erickcek et al.~\cite{erickcek}, is to assume that the perturbations in our Universe are modulated by a very-large-scale perturbation across our Universe. Such a perturbation cannot be the usual inflaton-generated curvature perturbation, because such a perturbation makes a large contribution to the CMB quadrupole --- the Grishchuk--Zel'dovich effect --- while if anything the observed CMB quadrupole is smaller than expected rather than larger. A suitable effect can however arise if the very-large-scale perturbation is generated by the curvaton mechanism \cite{erickcek,lyth}.

The nature of the modulating effect is one of cosmic variance. The modulating perturbation is presumably itself stochastic, corresponding to some power spectrum at very low wavenumbers $k \ll aH$, and the mean power spectrum on sub-horizon scales, averaged over all possible observers, is unmodulated. However a {\em typical}, rather than {\em average}, observer sees a stochastic short-scale spectrum superimposed on an effectively classical very-large-scale variation within that region, such that the power spectrum measured by a typical observer does feature the modulation.

\section{Modulation in an open universe}

A drawback of the modulation scenario as described in the literature thus far is the lack of an explanation for the existence of the modulating mode(s), other than a vague but non-calculable suggestion that they may somehow relate to the onset of inflation or the end of an eternal inflation stage \cite{erickcek}. This issue is substantially alleviated in the context of an open universe inflation model, both because the curvature scale sets a scale for super-horizon phenomena, and because open universes feature a new set of perturbation modes known as super-curvature modes that may be excited by inflation and can carry information about the pre-tunnelling vacuum state.

Super-curvature modes correspond to eigenmodes of the Laplacian with wavenumbers in the range $0<k^2<1$ when expressed in curvature units. These functions are not needed in order to expand a general radial function, for which the set with $k^2 \geq 1$ suffices to form a complete basis set, but they are nevertheless necessary in order to provide a general (scalar) random field. This was first discovered by mathematicians in the 1940s \cite{supercurv} and then deployed in open inflation models during the 1990s \cite{LythWosz}. Moreover, it was discovered that open inflation models typically excite a single one of those modes, whose amplitude can be found by matching the pre-tunnelling quantum fluctuations across the bubble wall into the open universe. If the vacuum energy density before tunnelling is much greater than that immediately afterwards, the amplitude of the super-curvature mode may be much higher than that of the subsequently generated spectrum of sub-curvature modes which provide the main contribution to CMB anisotropies on all scales \cite{openperts}. Hence, such models give exactly the type of power spectrum phenomenology required by the modulation scenario.

Existing open inflation perturbation calculations have focussed on adiabatic perturbations generated by the inflaton field. For the present purpose these are of no use; as in the flat case they will provide large contributions to the CMB quadrupole at amplitudes well below that needed to provide the modulation. Instead one needs to add a curvaton field $\sigma$ to an existing open inflation model. The simplest option for the curvaton is that it be massive but otherwise non-interacting during inflation (it will of course need to have decay channels after inflation to convert its perturbations into a curvature perturbation), i.e.\  with potential $V(\sigma) = \frac{1}{2} m_\sigma^2 \sigma^2$. Such a field adds three parameters to an open inflation model, being the mass $m_\sigma$, the `initial' value of the curvaton field $\sigma_*$ in our region of the Universe during inflation, and the curvaton decay rate that sets its conversion to normal matter.

The simplest option for the inflaton is a single-field model with a barrier and a flat region supporting slow-roll inflation after tunnelling, as proposed by Bucher et al.~\cite{BGT}. Given complete freedom to design such a potential it should be possible to create a working model, but we anticipate considerable fine-tuning may be needed to arrange for the super-curvature mode to be significantly enhanced over the normal spectrum (indeed, Ref.~\cite{LM} shows that considerable tuning is needed to get open inflation to work at all in such scenarios).

Instead, a more promising avenue is the two-field open inflation models introduced by Linde and Mezhlumian \cite{LM}, where the tunnelling is executed by one field, which we label $\psi$, and slow-roll inflation within the bubble is driven by a different field $\phi$. They consider two models:
\begin{eqnarray}
V_1 & = & \frac{1}{2} m^2 \phi^2 + W(\psi) \,; \label{LM1}  \\
V_2 & = & \frac{1}{2} g^2 \phi^2 \psi^2 + W(\psi)\,,
\end{eqnarray}
where in each case $W(\psi)$ is some tunnelling potential whose false vacuum is at $\psi = 0$ and whose precise shape does not appear crucial. 

The perturbations in these models were investigated in Ref.~\cite{openperts}, by propagating fluctuations in the $\phi$ field from the pre-tunnelling de Sitter phase across the bubble wall into the open universe created by the $\psi$ tunnelling (see also Ref.~\cite{juan} for an analysis of the perturbations in a one-field model). Their result is that the adiabatic super-curvature spectrum, for a nearly massless scalar, is a delta-function located at $(k^{\rm curv}_{\rm L})^2 \simeq 2m_{\rm F}^2/3H_{\rm F}^2$, where the wavenumber is specified in units of the curvature scale $k^{\rm curv} = 1/r_{\rm curv}$  and $m_{\rm F}$ is the field mass during the false vacuum stage. The amplitude is enhanced relative to the sub-curvature (continuum) spectrum by a factor $H^2_{{\rm F}}/H^2_{{\rm T}}$, where $H_{{\rm F}}$ and $H_{{\rm T}}$ are the Hubble parameters before and after the quantum tunnelling. This factor was envisaged to be large in the models introduced in Ref.~\cite{LM}.

The asymmetry $A$ is related to the non-gaussianity parameter $f_{\rm NL}$, given a single modulating mode with wavenumber $k_{\rm L}$, using \cite{lyth,NBF}  %
\begin{eqnarray}
\label{AitokPfNL}
|A| & = &  \frac{6}{5} |f_{\rm NL}| (k_{\rm L} x_{\rm ls}) P_{{\cal R}, {\rm L}}^{1/2}\,,\nonumber  \\
 & \simeq & \frac{18}{5} |f_{\rm NL}| |1-\Omega_{\rm total}|^{1/2} k^{{\rm curv}}_{\rm L} P_{{\cal R}, {\rm L}}^{1/2}\,,
\end{eqnarray} 
where $k_{\rm L}^{\rm curv}=k_{\rm L} x_{\rm ls}/(3 |1-\Omega_{\rm total}|^{1/2})$, $x_{\rm ls} \simeq 3 a^{-1} H^{-1}$ is the distance to last scattering, and $P_{{\cal R}, {\rm L}}$ is the super-curvature power spectrum [i.e.\ the coefficient of a delta-function contribution $\delta(\ln k - \ln k_{\rm L})$].

The highest achievable amplitude depends on the limits on the various terms on the right-hand side of Eq.~(\ref{AitokPfNL}). The non-gaussianity and departure from flatness are directly constrained by {\it Planck}.
The perturbation amplitude of the super-curvature mode is constrained by its contribution to the quadrupole via the Grishchuk--Zel'dovich effect. This has been computed in Ref.~\cite{openGZ} in the limit of flat geometry as
\begin{equation}\label{quadrupole}
6C_2^{\rm GZ} \simeq \frac{64}{625 \pi} (k^{\rm curv}_{\rm L})^2 P_{{\cal R}, {\rm L}} (1-\Omega_{\rm total})^2\,.
\end{equation}
The apparent strong dependence on $1-\Omega_{\rm total}$ arises because the expression considers a mode fixed in units of the curvature scale, and as $\Omega_{\rm total} \rightarrow 1$ that scale reaches arbitrarily long comoving wavelength. In the flat limit the quadrupole constraint is stronger than the octupole.
We adopt the same limit on such a contribution to the quadrupole as Ref.~\cite{erickcek}, $C_2^{\rm GZ} < 3.6\times 10^{-11}$, obtaining
\begin{equation}\label{quadrupole}
k^{\rm curv}_{\rm L} P^{1/2}_{{\cal R}, {\rm L}} |1-\Omega_{\rm total}| < 8 \times 10^{-5} \,.
\end{equation}
This implies
\begin{equation}
\label{result}
|A| < 3 \times 10^{-4} |f_{\rm NL}| |1-\Omega_{\rm total}|^{-1/2}\,.
\end{equation}

It appears from this expression that an arbitrarily large asymmetry could be obtained in the flat limit, but that is not the case due to the constraints that $k_{\rm L}^{\rm curv} <1$ to be super-curvature, and $P^{1/2}_{{\cal R}, {\rm L}} < 1$ for perturbation theory to apply (Lyth argues for a stronger limit $P^{1/2}_{{\cal R}, {\rm L}} < 1/|f_{\rm NL}|$ \cite{lyth}). These require $1-\Omega_{\rm total}$ to be greater than $8 \times 10^{-5}$ to saturate Eq.~(\ref{quadrupole}). Maximizing Eq.~(\ref{result}) under these constraints combined with Eq.~(\ref{quadrupole}) yields a maximum achievable asymmetry of
\begin{equation}
|A| < 0.03 |f_{\rm NL}|
\end{equation}
when these constraints are simultaneously saturated. This is very similar to results that have been obtained in the flat case, indicating that the flat-space limit is achieved smoothly giving a finite maximum asymmetry.

Equation (\ref{result}) shows that the required asymmetry can indeed be generated by a super-curvature perturbation, provided $|f_{\rm NL}|$ is not far from present limits. Interestingly, to maximize the asymmetry the curvature should be as small as allowed by the constraint for the super-curvature mode to exist, with viable scenarios requiring roughly $8 \times 10^{-5} < 1-\Omega_{\rm total} < 10^{-3}$ if $|f_{\rm NL}| \lesssim 10$.

We can convert these into bounds on the curvaton mass $m_{\rm F}$ and tunneling ratio $H_{\rm F}/H_{\rm T}$. In fact $H_{\rm F}$ cancels in the formula for $k_{\rm L}^{\rm curv} P_{{\cal R},{\rm L}}^{1/2}$, and  using $P_{{\cal R}, {\rm L}} \simeq (H_{\rm F}^2/H_{\rm T}^2)  P_{\cal R}$ taking into account the observed continuum spectrum normalization $P_{\cal R}^{1/2} \simeq 5 \times 10^{-5}$, we find
\begin{equation}
\frac{m_{\rm F}}{H_{\rm T}}  |1-\Omega_0| \lesssim 2 \,.
\end{equation}
The maximum achievable asymmetry remains as above, realized when this inequality is saturated. Typical parameters would be $1-\Omega_0 \simeq 10^{-4}$ and $m_{\rm F} \simeq 10^4 H_{\rm T}$, giving $|A| \simeq 0.01 |f_{\rm NL}|$. The parameter $H_{\rm F}$ is not directly constrained but consistency of the scenario requires it to lie in the range $(2/3) m_{\rm F}^2 < H_{\rm F}^2 < 4 \times 10^{8} H_{\rm T}^2$.

\section{A specific implementation}

As a proof of concept, we assemble these ideas into a particular model, which, while still a toy model, contains enough ingredients to generate the desired outcome. The overall potential is
\begin{equation}
\label{ourmodel}
{\cal L}  =  W(\psi) + \frac{1}{2} m_\phi^2 \phi^2 
 + \frac{1}{2} m_\sigma^2 \sigma^2 + \frac{1}{2} g^2 \psi^2 \sigma^2 \,,
\end{equation}
where all fields can be taken to be canonically normalized. The first two terms give exactly the Linde--Mezhlumian model \cite{LM}, where $W(\psi)$ is a tunnelling potential, though importantly we will define the true vacuum to be at $\psi = 0$ and place the false vacuum at $\psi_{{\rm F}}$. The $\phi$ field is responsible for inflation after tunnelling. The latter two terms contain the curvaton field $\sigma$, where we have included a direct coupling to the tunnelling field $\psi$ reminiscent of the second Linde--Mezhlumian model. The purpose of this coupling term is to permit the curvaton mass to change during the tunnelling. After tunnelling, the model effectively reduces to the simplest curvaton scenario of two massive non-interacting fields, as studied in Ref.~\cite{BL}. Presuming negligible inflaton perturbations, for the continuum spectrum this model predicts a spectral index $n_{\rm s} \simeq 0.98$ and a small tensor-to-scalar ratio, which is an acceptable fit to {\it Planck} data \cite{planckxxii}.

In our Lagrangian the $\sigma$ field appears in just the same way as $\phi$ in the first Linde--Mezhlumian model, Eq.~(\ref{LM1}), and hence the field perturbations of $\sigma$ have the same form as computed in Ref.~\cite{openperts}. This gives a curvaton spectrum featuring a sharp super-curvature spike and a nearly scale-invariant sub-curvature spectrum, as required for a successful modulation scenario. To avoid a similar spike in the inflaton spectrum, either the inflaton mass could be larger than $H_{\rm F}$, in which case the super-curvature mode would not exist at all for that field, or much less than the curvaton mass so that its wavenumber is much smaller than that of the curvaton spike.

The wavenumber and amplitude of the super-curvature mode are determined by the ratios $(m_{\sigma}^2+ g^2 \psi_{\rm F}^2)/H_{\rm F}^2$ and $H_{\rm F}^2/H_{\rm T}^2$ respectively; suitable values of each can be obtained by choice of the tunnelling potential shape. The $\phi$ field then drives inflation and its mass may be chosen to make its contribution to the curvature perturbation negligible. The curvaton decays once inflation is over, generating the curvature perturbation; the appropriate decay timescale can be fixed by choice of the curvaton decay constant which does not affect other observables. As the curvaton itself must not drive a period of inflation, it may be necessary to suppress its mass after tunnelling, hence our inclusion of a coupling to the tunnelling field which generates a large mass before tunnelling that reverts to a small `bare' mass $m_\sigma$ afterwards. More detailed calculations are required to demonstrate whether this term is really necessary, as a small curvaton mean value $\sigma_*$ may already ensure this condition.

Finally, we note that limits on power asymmetry from quasars \cite{hirata} and the small-angle CMB may require $A(k)$ to be scale dependent. We have not tried to address this but the requirement is no different from the flat case; proposals include isocurvature perturbations \cite{erickcek2} or a scale-dependent $f_{\rm NL}$ as in axion curvaton models \cite{lyth,axioncurvaton}.

We conclude that a suitably-constructed model of the type we have described could explain the origin of the large-scale modulating perturbation and connect it to curvature-scale effects in a marginally-open universe. Ours is the first proposal of a model which permits a complete first-principles calculation of a perturbation spectrum including a large-scale modulation effect of the observed amplitude.

\vspace*{5pt}

\begin{acknowledgments}
A.R.L\ was supported by the Science and Technology
Facilities Council [grant number ST/K006606/1], and M.C.\ by EU FP7 grant PIIF-GA-2011-300606 and FCT (Portugal).  This work was initiated during a visit to the Kavli Institute for Theoretical Physics under the programme `Primordial Cosmology', supported in part by the National Science Foundation under Grant No.\ NSF PHY11-25915. We thank Chris Hirata, Nemanja Kaloper, Antony Lewis, David Wands, and especially David Lyth for discussions.
\end{acknowledgments}


\end{document}